\documentclass[twocolumn,reprint,superscriptaddress,
 amsmath,amssymb,
 aps,prx,
]{revtex4-1}

\usepackage{graphicx}
\usepackage{dcolumn}
\usepackage{bm}
\usepackage{hyperref}
\usepackage{natbib}

\begin{document}

\preprint{APS/123-QED}

\title{Measurement of a $^7$Li tune-out wavelength\\by phase-patterned atom interferometry}

\author{Eric Copenhaver}
\email{eric.copenhaver@berkeley.edu}
\author{Kayleigh Cassella}
\affiliation{Department of Physics, University of California, Berkeley, Berkeley, CA 94720, USA}
\author{Robert Berghaus}
\affiliation{Fachbereich Physik, Technische Universit\"{a}t Darmstadt, Hochschulstra{\ss}e 12, 64289 Darmstadt, Germany}
\author{Holger M{\"u}ller}
\altaffiliation{Lawrence Berkeley National Laboratory, Berkeley, CA 94720, USA}
\affiliation{Department of Physics, University of California, Berkeley, Berkeley, CA 94720, USA}


\date{\today}

\begin{abstract}
Atom interferometers typically use the total populations the interferometer's output ports as the signal, but finer spatial structure can contain useful information. We pattern a matter-wave phase profile onto an atomic sample. An interferometer translates the phase into a measurable pattern in the atomic density that we use perform the first direct precision measurement of the $^7$Li tune-out wavelength near 671 nm. Expressed as a detuning from the $|2S_{1/2},F{=}2\rangle{\rightarrow}|2P_{1/2},F'{=}2\rangle$ transition, we find 3329.3(1.4) MHz for the tensor-shifted tune out of the $|2S_{1/2},F{=}2,m_F{=}0\rangle$ state with $\sigma^\pm$ light polarization and 3310.1(4.9) MHz for the tune out of the the scalar polarizability. This technique may be generalized for directly sensing spatially varying phase profiles.

\end{abstract}

\maketitle


\section{\label{sec:intro}Introduction}

In atom interferometers, phase differences between matter waves propagating on separated paths translate into measurable population differences at the output ports \cite{CroninReview}. The phase difference is typically uniform across the sample \cite{TinoG,TinoG1,KasevichG,FountainAlpha,BirabenAlpha,Gupta,PaulDE,MattDE,WuhanEEP,DennisEEP,InertialFrame,spinEEP,BonninEEP,TinoEEP} or has a constant gradient \cite{PSR,SpacetimeCurvature,Muntinga,CloseSeparation}. 
Detection methods tend to average out intricate spatial phase patterns that may be introduced during the interferometer. Imaging the atomic cloud accesses these patterns \cite{EdClock} and provides additional information that could be used for for atom lithography \cite{lithography} or to detect spatially-varying fields such as magnetic fields \cite{sqcram}, gravity gradients \cite{SpacetimeCurvature}, and thermal radiation \cite{blackbody}.

\begin{figure}
\includegraphics[width=0.47\textwidth]{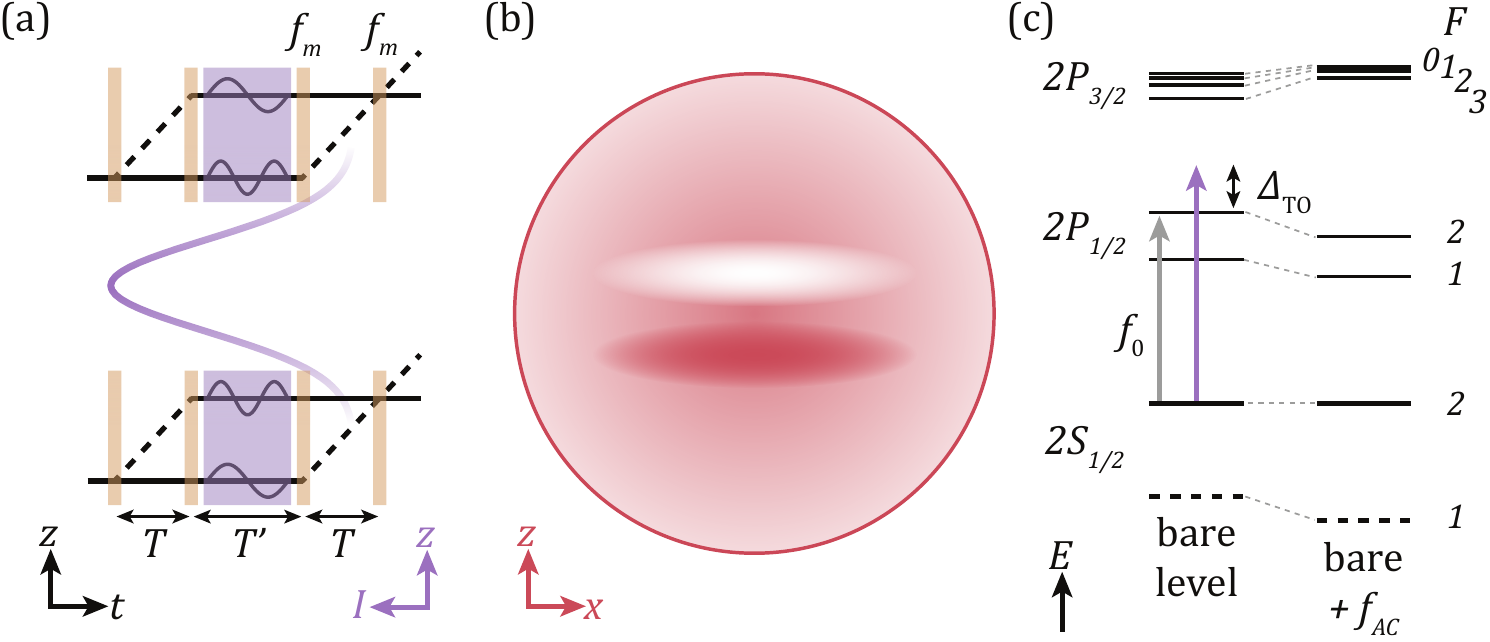}
\caption{\label{fig:principle} Principle of the experiment. (a) Solid (dashed) black lines indicate atom interferometer arms in $|F{=}2,m_F{=}0\rangle$ ($|F{=}1,m_F{=}0\rangle$). A Stark-shifting beam, with a spatial intensity profile (purple curve), focuses to a size smaller than the atomic sample. 
Gold vertical rectangles represent Raman pulses and the central purple rectangle represents the pulse of the Stark-shifting beam. The waves symbolize the phase shift accrued by the matter waves during the Stark pulse. Whichever arm of the interferometer is closer to the Stark beam center accrues a more dramatic phase shift. (b) Schematic of the atomic density that the interferometer outputs into $F{=}1$. Atoms on opposite sides of the beam center accrue opposite phase differences, which the interferometer translates into population differences. (c) Atomic energy levels are perturbed by the presence of the light field. Only when the Stark-shifting laser (purple arrow) is tuned to the tune-out wavelength, $\Delta_{\rm TO}$ away from the $|2S_{1/2},F{=}2\rangle{\rightarrow}|2P_{1/2},F'{=2}\rangle$ transition at optical frequency $f_0$, the polarizability vanishes and the perturbation on $|F{=}2,m_F{=}0\rangle$ is 0.}
\end{figure}

Here, we propose and demonstrate phase-patterned atom interferometry, where the signal source imprints a spatial phase pattern onto an atom interferometer (Fig. \ref{fig:principle}). Image analysis extracts the resulting population pattern despite a signal-to-noise ratio well below unity for any single image. We implement this technique to measure the tune-out (TO) wavelength of $^7$Li near 671\,nm, the wavelength where the ground state's polarizability $\alpha$, or AC Stark shift $f_{AC}$, vanishes. TO wavelengths are important in fundamental and applied physics, offering a versatile tool in quantum state engineering \cite{Arora}. Since TO wavelengths are unique to a specific state, they can be used to create species- and state-dependent potentials \cite{WideraMixture,CataniImpurity,Catani,RaselLaunch,Daley} and for minimizing measurement backaction \cite{FaradayProbe}. Precision measurements of TO wavelengths \cite{LamporesiFis1,WideraFis1,SackettMeasurement,SackettErratum,LamporesiFis1,WideraFis1,CroninTO,RaisaTO,DyTO,HeTO} may also be used to test all-order atomic theory \cite{Mitroy,Arora,Kien} and QED \cite{ZhangQED,Puchalski}. \emph{Ab initio} calculations in simple atoms with three or fewer electrons admit explicit accounting of electron-electron correlations \cite{Pipin,Tang,Bromley,Tang2,ZhangHe}. Comparing theory and experiment for lithium serves to benchmark approximation methods applicable to heavier atoms \cite{MariannaLi}. This makes lithium a strong candidate for a precision polarizability reference species \cite{Ravensbergen}. Here, we present the first direct measurement of a TO wavelength in lithium.


We implement phase-patterned interferometry in our TO measurement by focusing a laser beam to a size smaller than the atomic sample. The beam introduces AC Stark shift gradients that are opposite on opposite sides of its center. An interferometer translates the opposite gradients into measurable population differences (Fig. \ref{fig:principle}) that are proportional to the polarizability. At the TO wavelength, the coherent effect of the beam disappears. 




\section{\label{sec:expt}Experiment}

The experiment begins with $2{\times} 10^7$ $^7$Li atoms in a Magneto-Optical Trap (MOT). Optical Pumping (OP) prepares atoms in the $|F{=}2,m_F{=}0\rangle$ ground state \cite{Cassella}.

Four stimulated Raman $\pi/2$ pulses driven by two counter-propagating laser beams with a frequency difference near the $2S_{1/2}$ hyperfine splitting drive the interferometer along the $z$ axis. The first pulse separates the matter waves for $T{=}53\,\mu$s ($\Delta z{=}9\,\mu$m), while the second pulse brings the arms back into $|F{=}2,m_F{=}0\rangle$ for measuring TO. 

The phase-patterning pulse (purple in Fig. \ref{fig:principle}) addresses the interferometer for $\tau{=}100\,\mu$s between the second and third Raman pulses, propagating along the $\hat{y}$ imaging axis. 
A complementary interferometer occupying $|F{=}1,m_F{=}0\rangle$ during $T'$ can also close, but its TO frequency is different by roughly the 800-MHz $|2S_{1/2},F{=}1\rangle{\rightarrow}|2S_{1/2},F{=}2\rangle$ hyperfine splitting. If the complementary interferometer closes, it contributes its own phase-patterned signal even at TO for $|F{=}2,m_F{=}0\rangle$ and leads to a systematic shift in the measurement. Pulsing MOT repump light during $T'{=}110\,\mu$s destroys the coherence of the complementary interferometer (Appendix \ref{sec:decay}).

The third and fourth Raman pulses bring the interferometer arms back together and interfere them, with a frequency difference modified by $f_m$. Tuning $f_m$ to $f_m^+{=}24\,$kHz ($f_m^-{=}34\,$kHz) adds a bias to the phase difference between the interferometer arms. 
Any additional phase shift, like the Stark shift induced by the phase-patterning beam, produces changes to the atomic population output into $F{=}1$, the state detected by absorption imaging (see Fig. \ref{fig:decay}(d)). The population pattern imprinted by the Stark-shifting beam reverses upon tuning $f_m$ between $f_m^\pm$.

During the Stark-shifting laser pulse, each arm of the interferometer accrues a matter-wave phase $2\pi f_{AC}(\Delta_L,x,z)\tau$ according to the local AC Stark shift (where $\Delta_L$ is the laser's detuning from the $|2S_{1/2},F{=}2\rangle{\rightarrow}|2P_{1/2},F'{=}2\rangle$ transition). $f_{AC}$ can be expressed as a product of the frequency-dependent polarizability and the light intensity, varying linearly in small differences between $\Delta_L$ and the TO detuning $\Delta_{\rm TO}$ and proportionally to the local intensity $I(x,z){\propto}\exp(-2x^2/w_x^2)\exp(-2z^2/w_z^2)$ (where $x$ and $z$ are coordinates perpendicular to the beam axis and $w_z{\approx} 150\,\mu$m and $w_x{\approx} 600\,\mu$m describe the corresponding $1/e^2$-intensity ``waists'' for the anamorphic beam). The phase difference measured by the interferometer is proportional to the intensity difference between the arms. Therefore, the phase difference is linearly proportional to the local intensity gradient $dI(x,z)/dz$ and the separation between the arms, for small distances $\Delta z{\ll} w_z$ along the beam profile. This phase difference accrues over the interaction time $\tau$ as,
\begin{equation} 
\Delta\phi(\Delta_L,x,z)\propto \eta \frac{dI(x,z)}{dz}\Delta z(\Delta_{\rm TO}-\Delta_L)\tau, 
\label{eq:phase}
\end{equation}
where $\eta{=}{\pm}1$ parameterizes the sign of the phase sensitivity as chosen by biasing the interferometer phase via $f_m^\pm$. The intensity profile $I(x,z)$ directly patterns a phase profile onto the cloud, which is converted into a pattern in the final atomic density distribution. For the small phase differences induced in our experiment with respect to the bias points, the population difference introduced by the interferometer is proportional to $\sin(\Delta\phi){\approx}\Delta\phi$. The resulting dipole-shaped pattern reverses sign when the laser is tuned to the opposite side of TO or when the sign of $\eta$ is reversed by tuning $f_m^{\pm}$.

\begin{figure}
\includegraphics[width=0.42\textwidth]{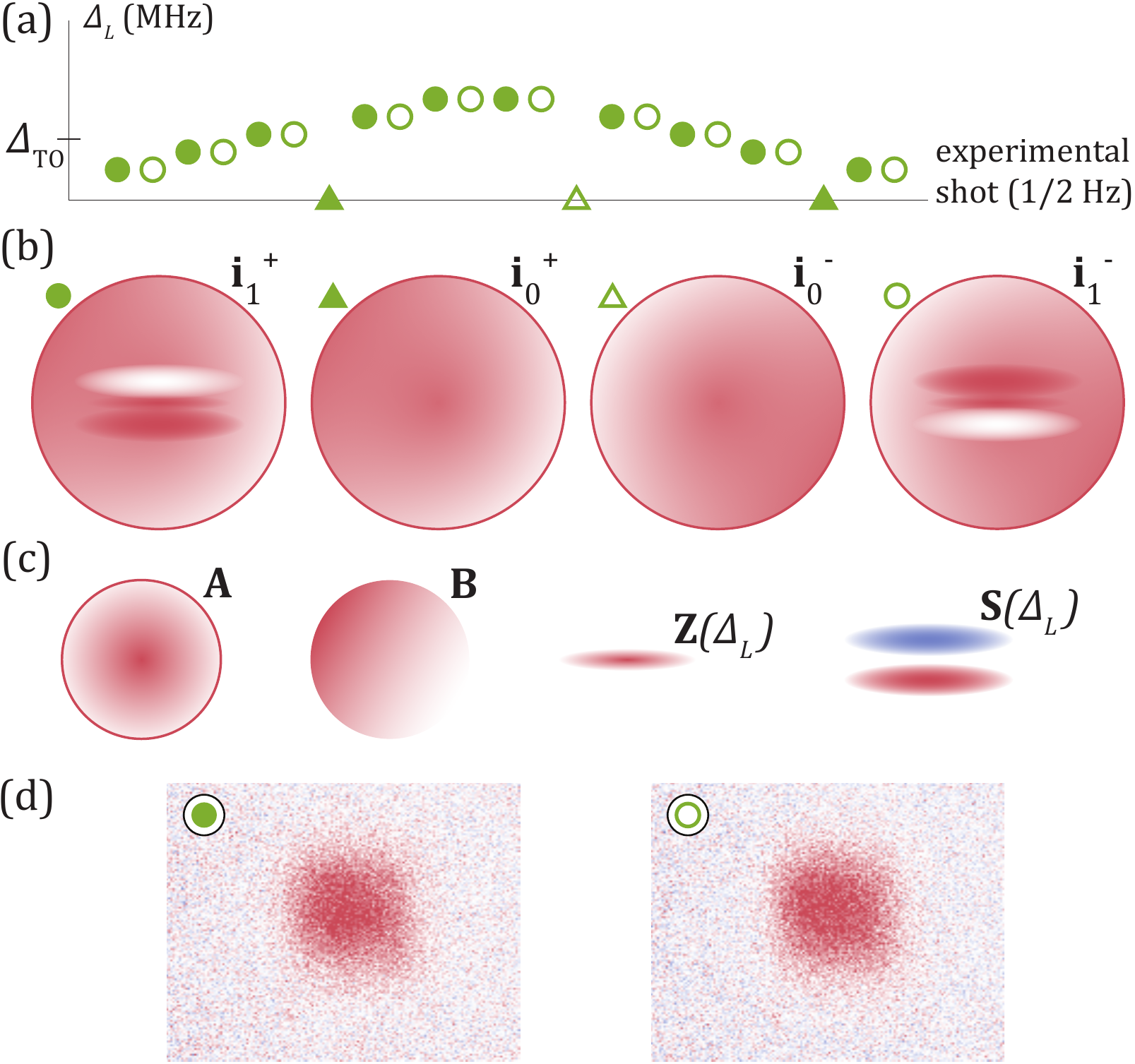}
\caption{\label{fig:analysis} Image processing. (a) As the Stark-shifting laser steps back and forth across $\Delta_{\rm TO}$, $f_m$ alternates between positive phase sensitivity with $f_m^+$ (filled circles) and negative phase sensitivity with $f_m^-$ (open circles). Shots without Stark pulses interrupt the pulsed shots at regular intervals, also with alternating positive phase sensitivity with $f_m^+$ (filled triangles) and negative phase sensitivity $f_m^-$ (open triangles). (b) Each image type exhibits features according to its sensitivity and Stark pulse state. (c) The images in (b) can be expressed as linear combinations of four features: the underlying atomic density profile ${\bf A}$, any population gradients ${\bf B}$ that arise from the interferometer independently of the Stark pulse, single-photon scattering from the Stark pulse ${\bf Z}(\Delta_L)$, and the signal of interest arising from coherent interaction with the Stark pulse ${\bf S}(\Delta_L)$. (d) These two raw images, one with each sensitivity, exhibit the effect of the Stark pulse with the highest single-shot sensitivity we achieve, i.e. with maximal $|\Delta_{\rm TO}{-}\Delta_L|$. The left (right) image has positive (negative) phase sensitivity. The signal pattern is subtle at best.}
\end{figure}

The small phase shifts imprint patterns with a signal-to-noise ratio too low to identify the pattern in any single image. Processing four types of alternating experimental shots extracts the interferometer signal from the Stark-shifting beam (Fig. \ref{fig:analysis}). Bold-face type denotes a two-dimensional image of pixel intensities in the $x{-}z$ imaging plane, ${\bf i}{=}i(x,z)$. Two shots at fixed Stark-shifting laser wavelength alternate in phase sensitivity (${\bf i}_1^\pm(\Delta_L)$, where the 1 indicates the presence of the Stark-shifting pulse and the $\pm$ represents the sign of $f_m^\pm$). To control the wavelength of the Stark-shifting laser, a phase-locked loop stabilizes an optical beat note ${\sim}$5-10 GHz below our master laser, which is locked near the $|2S_{1/2}\rangle{\rightarrow}|2P_{3/2}\rangle$ crossover resonance (Appendix \ref{sec:lock}). 
The Stark-shifting laser pulse is blocked after every tenth shot and the phase sensitivity also alternates for these unpulsed shots (${\bf i}^\pm_0$, where the 0 indicates the absence of the Stark pulse).

\begin{figure}
\includegraphics[width=0.47\textwidth]{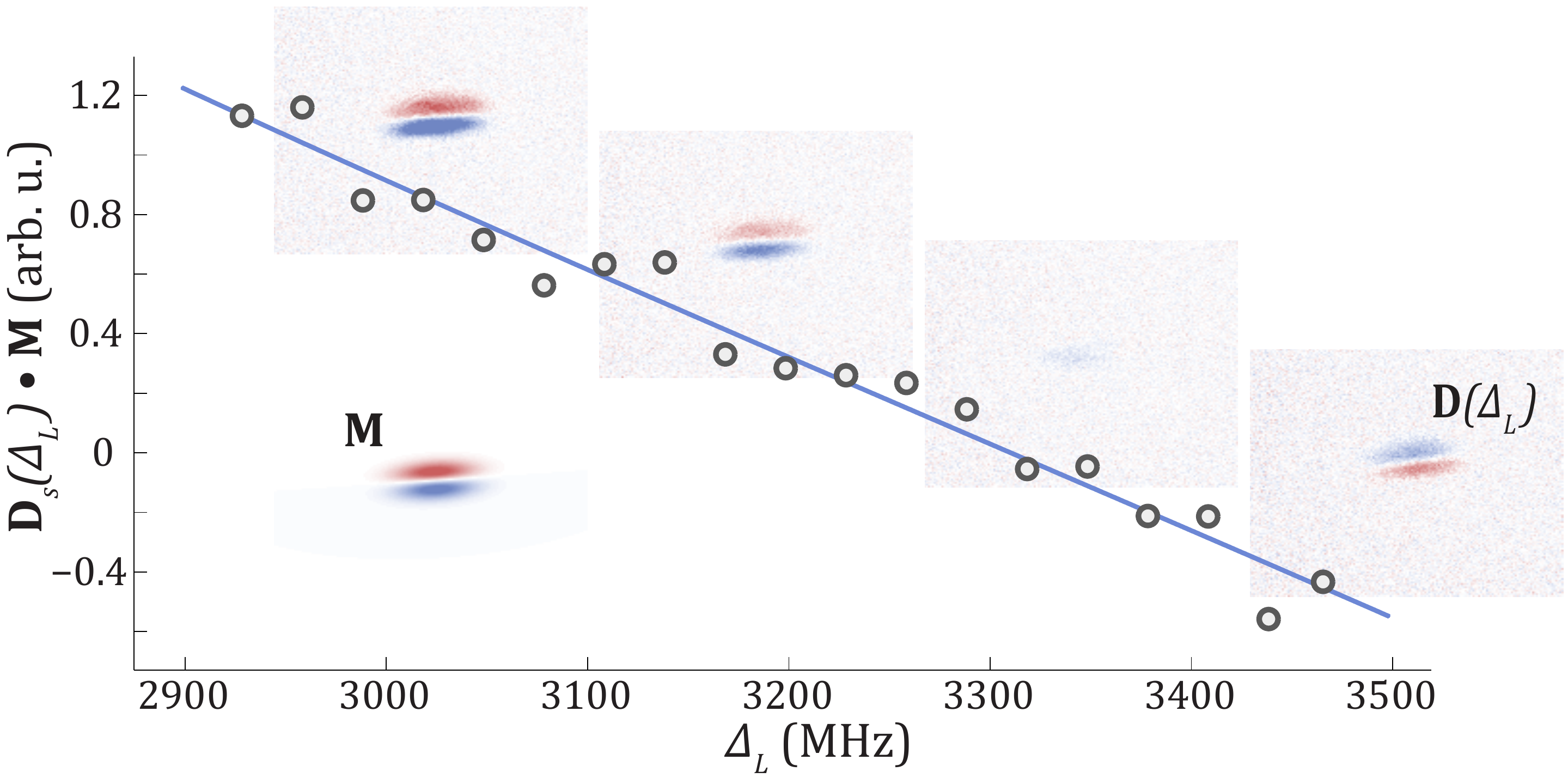}
\caption{\label{fig:DIms} Results of image processing. Projections ${\bf D}_s(\Delta_L)\cdot{\bf M}$ for a single subset (gray data points) cross through zero at $\Delta_{{\rm TO},s}$ (fit in blue). Insets show a selection of the ${\bf D}(\Delta_L)$ at different Stark laser frequencies (averaged over all subsets for clarity). ${\bf M}$ fits the average ${\bf D}(\Delta_L)$ with the highest contrast.}
\end{figure}

\begin{figure*}[t!]
\centering
\includegraphics[width=0.97\textwidth]{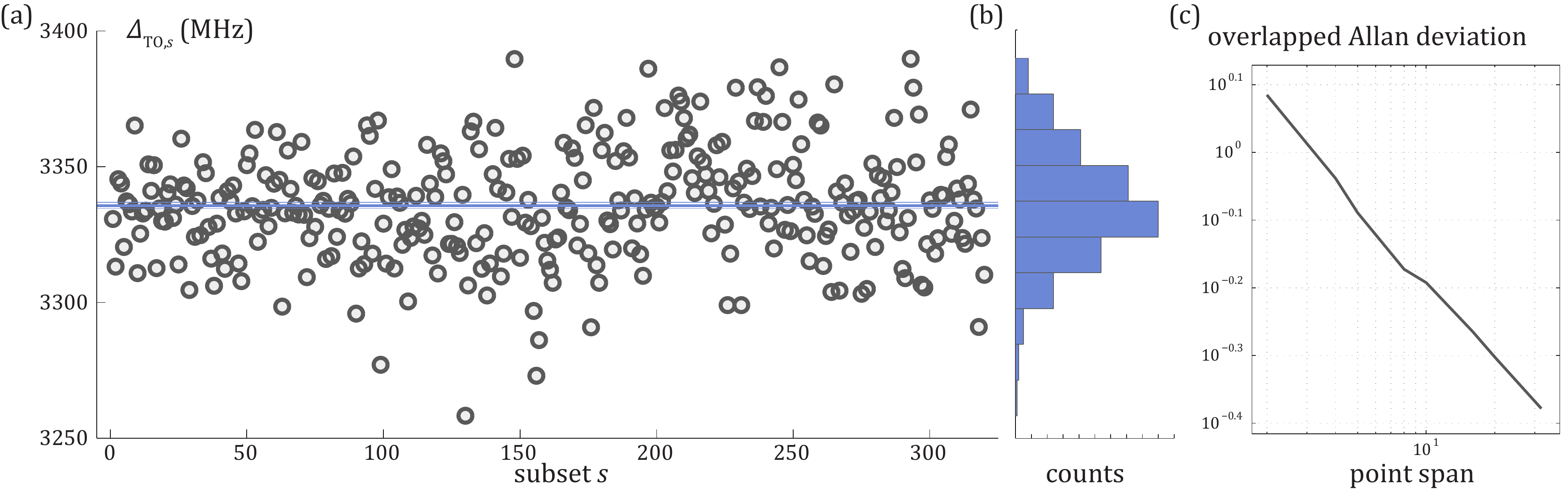}
\caption{\label{fig:stats} Statistics of image analysis results. (a) The fit result $\Delta_{{\rm TO},s}$ from each subset. Each data point is the zero-crossing from a plot like Fig. \ref{fig:DIms}. (b) A histogram of the results looks reasonably Gaussian, justifying using the standard error as an uncertainty metric. (c) The overlapped Allan deviation of the points in (a) shows the error integrating down with a power of -0.4.}
\end{figure*}

We isolate the pattern of interest by linearly combining averages of these image types, as outlined explicitly in Appendix \ref{sec:analysis}. Subtracting the averaged unpulsed shots from the averaged pulsed shots for each sensitivity 
\begin{equation}
{\bf R}^\pm(\Delta_L)={\bf i}_1^\pm(\Delta_L)-{\bf i}_0^\pm
\end{equation}
reveals the effect of the Stark pulse. Taking the difference between the resulting residual images 
\begin{equation}
{\bf D}(\Delta_L)={\bf R}^+(\Delta_L)-{\bf R}^-(\Delta_L)
\label{eq:firstD}
\end{equation} 
cancels the incoherent response from the pulse (single-photon scattering, ${\bf Z}(\Delta_L$)) and amplifies the sensitivity-dependent coherent response (AC Stark effect, ${\bf S}(\Delta_L$)), which vanishes when the laser is tuned to $\Delta_{\rm TO}$. We fit the ${\bf D}(\Delta_L)$ with the highest signal-to-noise ratio to a model ${\bf M}$ that follows Eq. (\ref{eq:phase}) for an anamorphic Gaussian beam. Projections of the difference images onto ${\bf M}$ cross through zero at TO (Fig. \ref{fig:DIms}). We perform this analysis on each of 320 subsets $s$ from the full data set, providing estimates of TO $\Delta_{{\rm TO},s}$ that bin to a Gaussian distribution and integrate down with nearly the square root of the integration time (a power of -0.4) to 1.2-MHz precision (Fig. \ref{fig:stats}). We anticipate this two-dimensional, projective analysis to be the most generally applicable approach for phase-patterned atom interferometry.

\section{Systematic effects}

\subsection{Spectroscopy}
The detunings $\Delta_L$ and $\Delta_{\rm TO}$ are referenced to spectroscopy of the cold atomic sample itself. This referencing allows us to bypass a ${\sim}10$-MHz inaccuracy in the master laser spectroscopy and to calibrate an important Doppler systematic. The Stark laser performs spectroscopy of the optically pumped cold-atom sample on the $|2S_{1/2},F{=}2\rangle{\rightarrow}|2P_{1/2},F'{=}2\rangle$ transition along two axes: $\hat{z}$ and the axis we use to measure TO. Spectroscopy along $\hat{z}$ in the imaging plane permits a calibration of the corresponding Doppler shift by measuring the launch speed along $\hat{z}$ in time-of-flight images. Correcting for the corresponding Doppler shift establishes the optical reference frequency $f_0$. 
The sample's center-of-mass velocity may also have a component along the Stark beam axis used to measure TO during interferometry (${\approx}\hat{y}$) that cannot be measured with our time-of-flight imaging. To quantify this, we perform spectroscopy with the Stark-shifting beam as it propagates along the TO axis, then compare the result to the Doppler-corrected $f_0$ from spectroscopy along $\hat{z}$. We attribute any discrepancy to the Doppler shift along that axis and correct the final TO measurement in Table \ref{tab:result}. Details of the spectroscopy analysis are presented in Appendix \ref{sec:spec}.

While our Stark-shifting laser has a suitably narrow linewidth of 1 MHz, it also emits Amplified Spontaneous Emission (ASE) \cite{ASE}. The broadband power spectrum spans ${\sim}100\,$nm and can contribute significant shifts in TO measurements. In Appendix \ref{sec:ase}, we present a measurement of the optical spectrum of the laser and use it to calculate a small systematic shift included in Table \ref{tab:result}.

\subsection{Polarization}
The polarizability of an atom in a particular hyperfine state $F$ and Zeeman sublevel $m_F$ can be decomposed into a scalar term ($\alpha^s$) and a pair of polarization-dependent vector and tensor terms ($\alpha^v$ and $\alpha^T$, respectively) \cite{Kien,Krutzik}.
\begin{align}
\alpha=&\alpha^s+C\frac{m_F}{2F}\alpha^v-D\frac{3m_F^2-F(F+1)}{2F(2F+1)}\alpha^T.
\label{eq:polarizability}
\end{align}
The factors $C{=}|e_{-1}|^2{-}|e_{+1}|^2$ and $D{=}1{-}3|e_0|^2$ depend on the circular component magnitudes of the light's polarization vector $\vec{e}$ ($\hat{e}_{\mp 1}{=}\hat{\sigma}^\pm$ and $\hat{e}_0{=}\hat{\pi}$). Because the scalar polarizability in Eq. (\ref{eq:polarizability}) dominates at most wavelengths and is independent of experimental geometry, the TO wavelength is conventionally defined as the wavelength at which specifically the scalar polarizability vanishes: $\alpha^s(\Delta^*_{\rm TO}){=}0$. We typically refer to TO in this paper more experimentally, as the condition at which the total polarizability is 0, given a specific state and light polarization: $\alpha(\Delta_{\rm TO}){=}0$. 

\begin{figure}[t]
\centering
\includegraphics[width=0.42\textwidth]{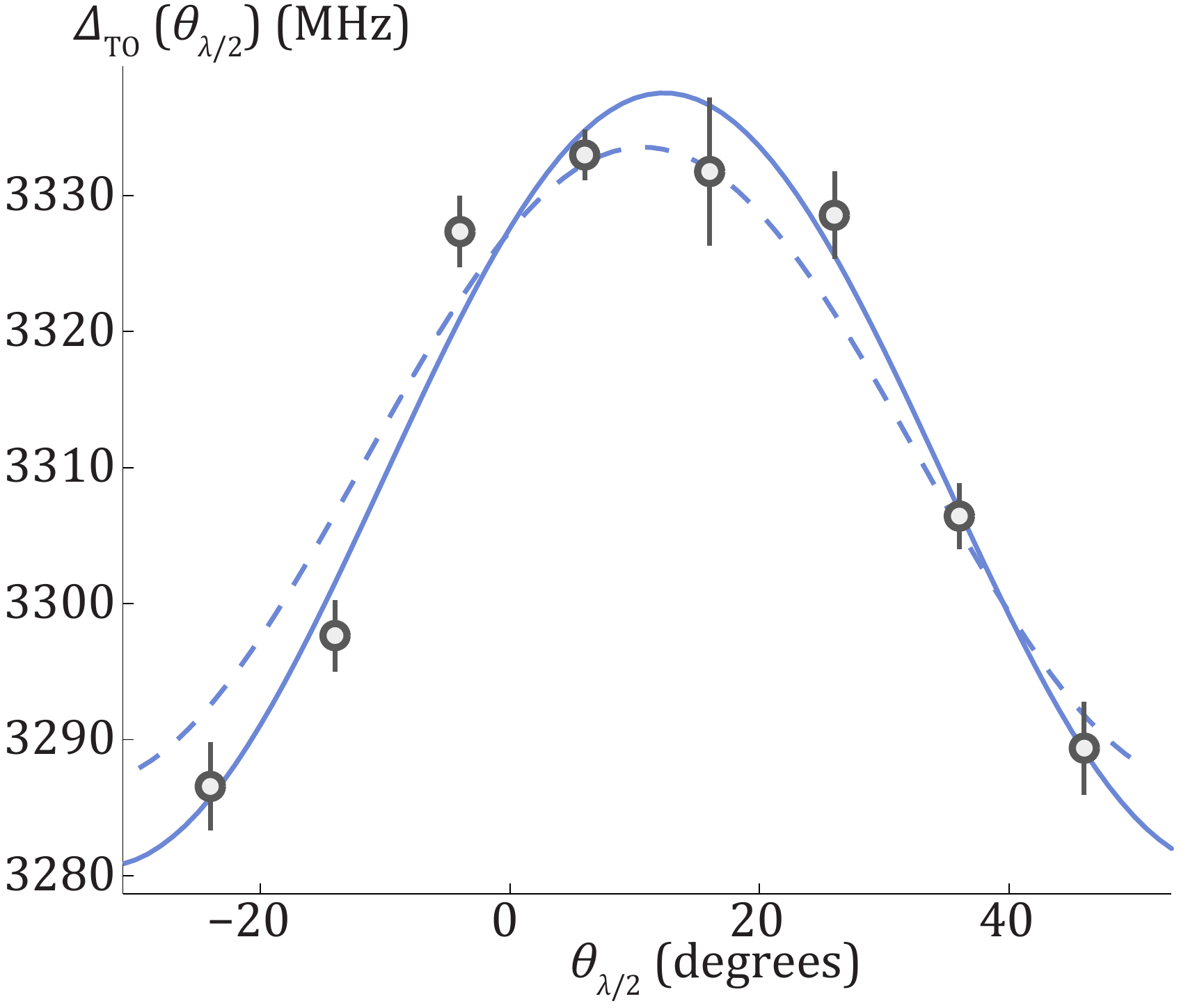}
\caption{\label{fig:tensor} The measured TO detuning varies with the polarization. The variation fits to a full amplitude of 56.9(4.7) MHz (solid line). For comparison, we show a fit constrained to the theoretical variation of 47 MHz (dashed line). 
The $\lambda/4$ wave plate angle is fixed in relationship to the $\lambda/2$ according to $(\theta_{\lambda/2}{-}6^\circ){=}2(\theta_{\lambda/4}-4^\circ)$. See Appendix \ref{sec:polarization} for details.}
\end{figure}

The tensor term in Eq. \ref{eq:polarizability} gives rise to a ${\sim}50$-MHz shift to TO between $\pi$ and $\sigma^\pm$ polarizations for $|F{=}2,m_F{=}0\rangle$. A Wollaston prism purifies the Stark beam's polarization and a pair of wave plates (one $\lambda/2$ and one $\lambda/4$) control the polarization. We use two methods to ensure that the polarization is a linear combination of $\sigma^\pm$, which contains no $\pi$ component for the final TO measurement (see Appendix \ref{sec:polarization}). We also scan out the full tensor shift by rotating the linear polarization through 8 values in Fig. \ref{fig:tensor}. 

\subsection{Single-photon scattering}
Our particular geometry permits further reduction of the two-dimensional signal into one dimension along the interferometer axis $\hat{z}$. The one-dimensional signal simplifies evaluation of a systematic effect from imperfect cancellation of the single-photon scattering pattern in ${\bf D}(\Delta_L)$. Any position offset between the remaining scattering pattern and the signal produces a systematic offset in the projections presented in Fig. \ref{fig:DIms} and the results they produce in Fig. \ref{fig:stats}. Single-photon scattering is particularly strong in this experiment due to the proximity of TO to the $|2S_{1/2}\rangle{\rightarrow}|2P_{1/2}\rangle$ and $|2S_{1/2}\rangle{\rightarrow}|2P_{3/2}\rangle$ transitions in Li. 

To generate a one-dimensional signal, we average ${\bf D}(\Delta_L)$ over all subsets and integrate the images perpendicular to the interferometer axis. The resulting trace for each $\Delta_L$ is a sum of the residual effect of scattering and the phase-patterned signal of interest. We fit each trace to a sum of those components with independent signal and scattering amplitudes and plot the amplitude of the signal portion of the fit. It crosses through zero at $\Delta_{\rm TO}$, the fit result presented in Table \ref{tab:result}.

\begin{table}
\caption{Systematic effects in measuring TO for $|2S_{1/2},F{=}2,m_F{=}0\rangle$ relative to the $|2S_{1/2},F{=}2\rangle{\rightarrow}|2P_{1/2},F'{=}2\rangle$ transition with $\sigma^\pm$ polarization. All frequencies are given in MHz.}
\begin{tabular}{|l|l|l|}\hline
Effect (relevant Appendix) & Correction & 1-$\sigma$ uncertainty \\
\hline\hline
Doppler shift (\ref{sec:spec}) & $+1.58$ & $0.06$ \\
Broadband laser emission (\ref{sec:ase}) & $-0.09$ & $0.1$ \\
$F{=}1$ interference (\ref{sec:decay}) & $-0.02$ & $0.04$ \\
$f_0$ Zeeman shift (\ref{sec:specZeeman}) & $-0.09$ & $0.02$ \\
$f_0$ statistical & & $0.05$ \\
Polarization impurity (\ref{sec:polarization}) & & $0.3$ \\
\hline
Total & +1.38 & 0.33 \\
\hline
\hline
One-dimensional fit (\ref{sec:scattering}) & 3327.95 & 1.40 \\
\hline
\hline
Final result & 3329.33 & 1.44 \\
\hline
\end{tabular}
\label{tab:result}
\end{table}

\section{Results}

Our measurement of TO for $|2S_{1/2},F{=}2,m_F{=}0\rangle$ in $^7$Li yields a result of $\Delta_{\rm TO}{=}$3329.3(1.4) MHz from the $|2S_{1/2},F{=}2\rangle{\rightarrow}|2P_{1/2},F'{=}2\rangle$ transition, a wavelength precision of 2.2 fm.

We compare our experimental result to the established atomic theory (Eq. \ref{eq:polarizability}) using a hyperfine basis \cite{Kien,Krutzik} to calculate the optical frequency that satisfies $\alpha(\Delta_{{\rm TO},th}){=}0$ for $|F{=}2,m_F{=}0\rangle$ and $\sigma^\pm$ polarization. One can obtain the same result by solving for the wavelength at which $f_{AC}$ vanishes, where the Stark shift is summed over each hyperfine transition whose coupling strength depends on a state- and polarization-dependent geometric (Clebsch-Gordan) coefficient. Theoretical matrix elements from \cite{MariannaLi} and experimental transition energies from \cite{Sansonetti} predict $\Delta_{{\rm TO},th}{=}$3323.5(1.3) MHz, in slight 3-$\sigma$ tension with our measurement. While our calculation neglects the effects of the core polarizability and states beyond the $n{=}2$ doublet, those contributions enter below our precision near the 0.1-kHz level in lithium. We also note that theory predicts the size of the polarization-dependent tensor shift to TO be 47 MHz, while we observe a 56.9(4.7) MHz modulation (a 2-$\sigma$ tension). 
Our measurement provides a less precise estimate of the scalar TO, dominated by our uncertainty in the polarization-dependent tensor shift. Calculating the tensor shift in Eq. (\ref{eq:polarizability}) as a function of polarization shows that the scalar TO sits 1/3 of the full tensor shift below the result for $\sigma^\pm$ polarization. Theory predicts the scalar polarizability for $|2S_{1/2},F{=}2\rangle$ to vanish at $\Delta_{{\rm TO},th}^*{=}$3308.1(1.3) MHz. Using our measured 56.9(4.7)-MHz tensor shift, we estimate scalar tune out for $|2S_{1/2},F{=}2\rangle$ to occur at $\Delta_{\rm TO}^*{=}$3310.6(4.9) MHz, in good agreement with theory. 

The comparison suggests that the tension between our measurement and theory might be fully described by a tension in the tensor shift. 
We may speculate about the possibility of an undetected polarization-related effect. For example, the polarization of the ASE pedestal also changes as we change the polarization of the Stark-shifting beam. If there were spectral asymmetries near the atomic resonances where our methods were unable to probe (see Appendix \ref{sec:ase}), then the ASE's coupling strength may also have varied with the polarization in the results plotted in Fig. \ref{fig:tensor}. 

\section{Discussion}

In summary, we pattern a phase profile into an atomic sample using a laser and an atom interferometer, applying the technique to measure the tune-out wavelength of $^7$Li's $|2S_{1/2},F{=}2,m_F{=}0\rangle$ state. Image analysis of modulated shots reads out the faint signal. The measurement references an atomic transition probed at the cold atomic sample, allowing for a calibration of the Doppler effect. 

These phase-patterning results establish a foundation for processing more complicated density distributions. Designer density profiles could be engineered by imprinting intensity gradients onto the Stark-shifting laser using a spatial light modulator. This could be combined with matter-wave lensing \cite{pklensing,Muntinga,svenlensing,Anciaux} to reduce feature sizes below the laser's diffraction limit.
It could even be performed in three dimensions in parallel \cite{3dai} to create nontrivial 3D density distributions. Extensions of our method may also aid in sensing fields that vary spatially. Blackbody \cite{blackbody,MSBBR,MSSrBBR} and lattice light shifts \cite{EdClock,MSlattice} in atomic clock outputs, for example, are emerging as a leading systematic in those experiments capable of searching for variation of the fundamental constants \cite{Peik}.


Tune-out measurements enjoy a fortunate circumstance in which current theoretical and experimental uncertainties are comparable. Light species like lithium offer particularly fruitful interplay between theory and experiment because of the high-accuracy calculational techniques available for few-electron species. \emph{Ab initio} calculations in atoms with few electrons can account for electron-electron correlations using the variational Hylleraas basis set \cite{Pipin,Tang,Bromley,Tang2,ZhangHe}. The Hylleraas calculations are expected to produce the most accurate results in lithium, where they serve as a benchmark for approximation methods applicable to heavier atoms \cite{MariannaLi}. Measurement of TO in metastable helium \cite{HeTO} has also inspired a rich interaction with theory \cite{ZhangQED,Drake}. Our tension with theory suggests that an independent measurement of TO in ultracold $^{6,7}$Li \cite{KasevichLi,Geiger,Dimitrova,HuletSolitons,Wright,salomonLi,Biswaroop,Omran,GreinerLattice} could add a valuable contribution to the dialogue between theory and experiment.

\begin{acknowledgements}
The authors are grateful the following scientists for useful discussions and help with the apparatus: Cass A. Sackett, Adam Fallon, Marianna Safronova, Osip Schwartz, Sara Campbell, G. Edward Marti, Victoria Xu, Matt Jaffe, Richard Parker, Zachary Pagel, Neil Goeckner-Wald, Chris Overstreet, Jason Hogan, Raisa Trubko, Michael Bromley, Swaantje J. Grunefeld, Yaron Hadad, Prabudhya Bhattacharyya, Thomas Mittiga, Andrew McNeely, and Satcher Hsieh. This work relies on a diverse and inclusive environment for all contributing scientists. This work is supported by the National Science Foundation under Grant No. 031510 and by the David and Lucile Packard Foundation.
\end{acknowledgements}

\appendix
\section{Experimental details}
\subsection{State preparation}

\begin{figure}
\includegraphics[width=0.25\textwidth]{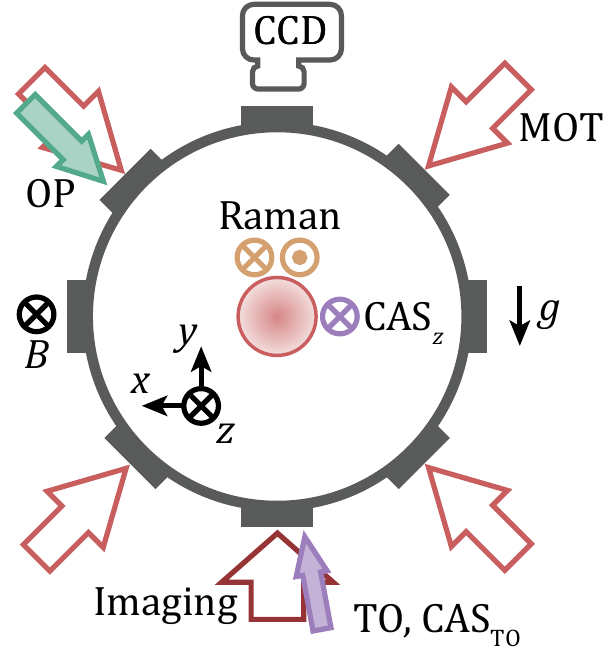}
\caption{\label{fig:axes} Description of experimental axes. The cloud is shown in red at the center of the gray vacuum chamber. The OP beam (green) propagates diagonally in the $x-y$ plane with linear polarization parallel to  the magnetic field $B$ that defines the $z$ axis. Counter-propagating Raman beams (gold) propagate close to $\hat{z}$ with orthogonal linear polarizations, one along $\hat{x}$ and one along $\hat{y}$. The imaging beam (dark red) propagates anti-parallel to gravity $g$ and produces absorption images of $F{=}1$ at the charge-coupled device (CCD) camera. Four MOT beams (red) lie in the $x{-}y$ plane with two more along $\hat{z}$ (not shown). The Stark laser (purple) probes TO along an axis as close to the imaging axis as possible during interferometry, with polarization in the $x{-}y$ plane close to $\hat{x}$. The Stark laser also performs cold-atom spectroscopy along the TO measurement axis (CAS$_{\rm TO}$) and along $\hat{z}$ in the imaging plane (CAS$_z$), both with polarization close to $\hat{x}$.}
\end{figure}

The experiment begins by laser cooling and trapping roughly $2{\times}10^7$ $^7$Li atoms in a MOT \cite{Cassella}. The high thermal speeds of the atoms near the Doppler temperature ($T_D{\approx}$140 ${\rm \mu K}$, $v_{th}{\approx}0.7$ m/s) demand hasty state preparation, interferometry, and imaging. After turning off the MOT magnetic quadrupole field, we wait about 2\,ms to allow the gradient to decay with ${\sim}1$-ms time constant; during this time, an optical molasses limits expansion of the sample. The residual field gradient of 0.5\,G/cm is small enough compared to the 1.3-G bias magnetic field to establish a homogeneous quantization axis that defines the $\hat{z}$ axis. After molasses, the sample's center-of-mass velocity $\vec{v}_l$ has a component of roughly 1.5\,m/s in the $x{-}z$ absorption imaging plane, likely due to the decay of the magnetic gradient in the presence of the nonzero bias field. Fig. \ref{fig:axes} shows a detailed summary of the axes.



To optically pump (OP) the atoms into the magnetically insensitive $|F{=}2,m_F{=}0\rangle$ state, we apply light tuned to the $|2S_{1/2},F{=}2\rangle{\rightarrow}|2P_{1/2},F'{=}2\rangle$ transition for 20\, ${\rm \mu s}$. The linearly polarized OP beam propagates diagonally in the $x{-}y$ plane and is retro-reflected. It must be purely $\pi$-polarized for efficient pumping, so we rotate the angle of the magnetic bias field using 3 axes of Helmholtz coils to optimize OP efficiency. MOT repump light repumps atoms decaying into $F{=}1$ during OP. Microwave spectroscopy shows that 80\% of the atoms exit OP in the $|F{=}2,m_F{=}0\rangle$ state. Roughly 16\% remain in $m_F{=}{-}2$, 4\% in $m_F{=}{+}2$, and no atoms are detectable in the $m_F{=}{\pm}1$ states or the $F{=}1$ manifold. The final atomic density distribution is approximately Gaussian with a waist of $w_a{\approx}500$ ${\rm \mu m}$, far larger than the interferometer arm separation.

\subsection{\label{sec:lock}Laser lock}
A master laser stabilizes the frequencies of the cooling and trapping lasers. It is an External Cavity Diode Laser (ECDL), frequency-stabilized (``locked") near the ground-state's crossover resonance between hyperfine states on the $|2S_{1/2}\rangle{\rightarrow} |2P_{3/2}\rangle$ transition using Modulation Transfer Spectroscopy (MTS) of a hot lithium spectroscopy cell. The MTS lock is stable to ${\sim}100\,$kHz, but is offset from the true crossover resonance by ${\sim}$10 MHz due in part to an asymmetric error signal. This offset precludes it from serving as the reference frequency from which we measure $\Delta_{\rm TO}$. Light from the master laser injection-locks the diode laser that generates light for the Raman beams driving the atom interferometer (Appendix \ref{sec:aiapp}).

The Stark-shifting laser that performs the phase patterning and cold-atom spectroscopy is an ECDL (Toptica, DLC DL PRO 670), which is offset-locked to the light used to drive Raman transitions. A phase lock feeds back to the Stark laser's current and grating angle for the offset lock. The frequency of the local oscillator (LO) in the offset lock is tuned around 2.5 GHz, which is either tripled to lock the laser near the TO wavelength during interferometry or quadrupled to perform cold-atom spectroscopy of the $|2S_{1/2}\rangle{\rightarrow} |2P_{1/2}\rangle$ transition, ${\sim}$10 GHz below the $|2S_{1/2}\rangle{\rightarrow}|2P_{3/2}\rangle$ transition. 
The laser spectrum ultimately inherits a FWHM Gaussian linewidth of 1 MHz. Its entry into the vacuum chamber is shuttered by a 180-MHz acousto-optical modulator (AOM), both for the Stark-shifting pulse during interferometry and for cold-atom spectroscopy along both axes described below. 

\subsection{Cold-atom spectroscopy}
\label{sec:spec}

The frequency reference provided by the master laser is not accurate enough to serve as a reference in the TO measurement. To establish a more accurate reference for our TO measurement relative to a specific transition, we perform cold-atom spectroscopy (CAS) of the $|2S_{1/2}\rangle{\rightarrow}|2P_{1/2}\rangle$ transition with the Stark-shifting laser on the optically pumped sample. To help calibrate Doppler effects arising from motion of the cold atoms, we perform CAS along the $z$ axis (CAS$_z$) and the axis we use to measure TO (CAS$_{\rm TO}$). The Stark-shifting laser lock and AOM shutter remain identical for CAS$_z$ and CAS$_{\rm TO}$; the only difference is a magnetic mirror that optionally redirects the beam to the $z$ axis from the TO axis.

We specify laser frequencies $f_L$ relative to the optical frequency of the master laser $f_{\rm MTS}$. The difference between the master laser and the Stark-shifting laser at the atoms is set by the LO frequency in the offset lock (quadrupled for CAS or tripled for interferometry) $\Delta f$ and some frequency offsets $f_{\rm off}$ introduced by the AOM shutter and phase lock.
\begin{equation}
f_L = f_{\rm MTS}-\Delta f+f_{\rm off}
\end{equation}
Taking differences between laser frequencies measured in this way cancels the common-mode $f_{\rm off}$ term (see Eq. \ref{eq:detuning}).

\begin{figure}[t]
\centering
\includegraphics[width=0.47\textwidth]{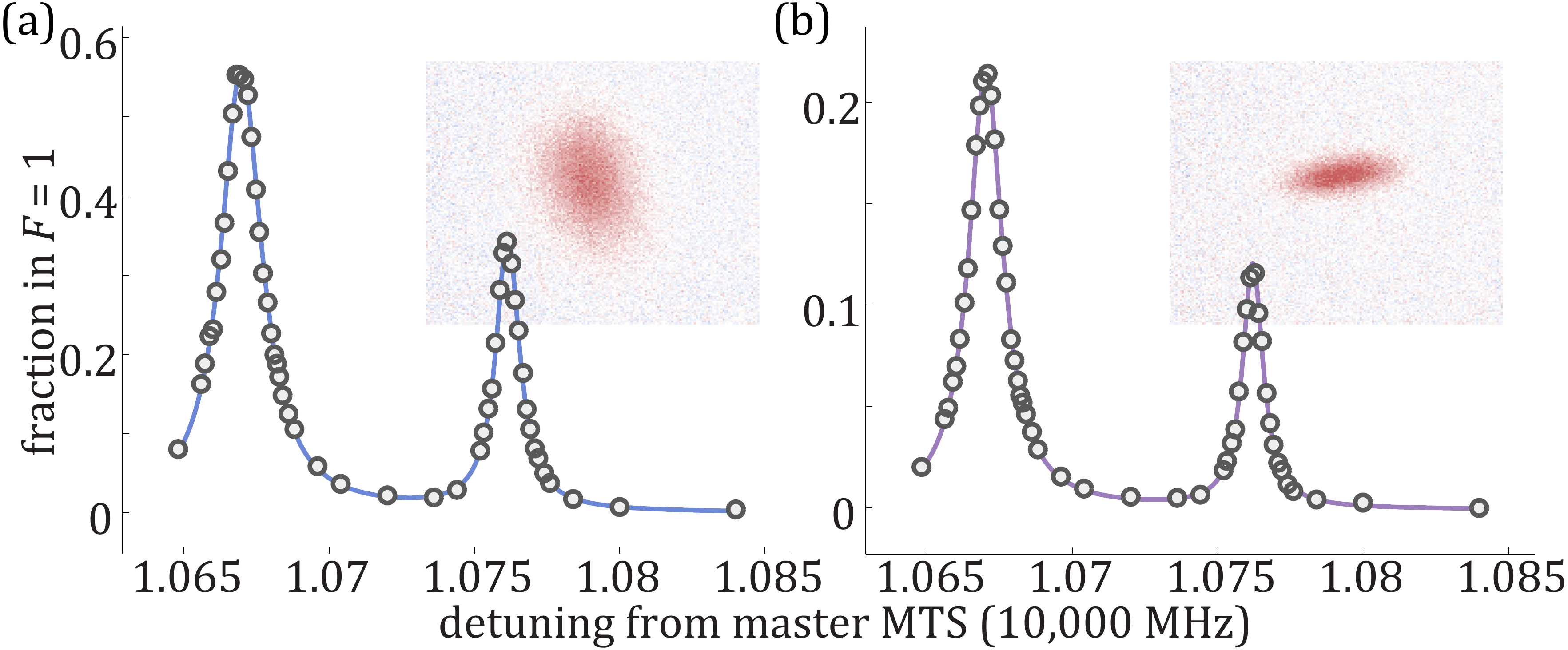}
\caption{\label{fig:cas} Cold-atom spectroscopy. (a) Spectroscopy along $\hat{z}$ in the imaging plane permits a calibration of the Doppler shift along that axis. The blue line is a fit to the sum of two Lorentzian peaks. The average fit result to the left peak serves as a reference for all detunings in this paper. (b) Spectroscopy along the TO axis helps calibrate any Doppler shift along that axis. The purple line is a fit to the sum of two Lorentzian peaks. Comparing the average fit results to the Doppler-corrected results in (a) provides an estimate of the Doppler shift along the Stark beam axis. Each inset shows an absorption image of the atomic population detected in $F{=}1$ when spectroscopy is performed along the corresponding axis.}
\end{figure}

The center-of-mass velocity $\vec{v}_l$ of the sample produces a Doppler shift and introduces an important Doppler systematic to the TO measurement performed along that axis. To calibrate the shift, we first perform CAS$_z$ on an axis in the imaging plane where we can measure the component of the launch speed along the spectroscopy axis $v_z$ by fitting the $z$-position of the cloud in time-of-flight absorption images. During CAS$_z$, the Stark-shifting laser propagates within $5^\circ$ of $\hat{z}$. Its linear polarization is roughly parallel to $\hat{x}$, i.e. $(\sigma^+{+}\sigma^-)/\sqrt{2}$. This beam has a $1/e^2$-intensity waist of 1.5 mm and power of $\lesssim$0.4 ${\rm mW}$. Scanning the LO frequency scans the detuning of the Stark-shifting laser with respect to the master MTS spectroscopy. As a function of the LO frequency near the $|2P_{1/2}\rangle{\rightarrow}|2P_{3/2}\rangle$ fine-structure splitting, the fraction of the atoms appearing in images of $F{=}1$ reveals two resolved peaks from the $|2S_{1/2}\rangle{\rightarrow} |2P_{1/2}\rangle$ transition (Fig. \ref{fig:cas}). The $|2S_{1/2},F{=}2\rangle{\rightarrow} |2P_{1/2},F'{=}2\rangle$ transition appears with lower LO frequency (higher laser frequency) at $\Delta f^{D1,22}_z$. The $|2S_{1/2},F{=}2\rangle{\rightarrow} |2P_{1/2},F'{=}1\rangle$ transition appears with higher LO frequency (lower laser frequency) at $\Delta f^{D1,21}_z$. We fit the results to the sum of two Lorentzians.

We correct the shifted CAS$_z$ resonance by an amount $v_z k^{D1,22}_z$, where $k^{D1,22}_z$ is the $z$-component of the wave vector of this light. 
\begin{equation}
\label{eq:DopplerCorrectedf0}
f_0 = f_{\rm MTS} - (\Delta f^{D1,22}_z+v_z k^{D1,22}_z) + f_{\rm off}.
\end{equation}
This constitutes a single measurement of the spectroscopic baseline. We interleave these spectroscopy measurements 14 times throughout the TO measurement campaign and quote optical frequencies as detunings relative to the average result $\langle \Delta f^{D1,22}_z\rangle{+}\langle v_z k^{D1,22}_z\rangle{=}10668.36(5)$ MHz.
\begin{equation}
\Delta_L = f_L-\langle f_0\rangle.
\label{eq:detuning}
\end{equation}

Following each CAS$_z$ measurement of $f_0$, we also perform CAS$_{\rm TO}$ on the axis along which the Stark laser propagates during the interferometry sequence, within $5^\circ$ of $\hat{y}$. The polarization of this beam lies in the $x{-}y$ plane and is a linear combination of $\sigma^\pm$ with roughly equal weights. We compare the average result along this axis $\langle \Delta f^{D1,22}_{\rm TO}\rangle$ to $\langle \Delta f^{D1,22}_z\rangle {+}\langle v_z k^{D1,22}_z\rangle$ and attribute any discrepancy to the Doppler effect along the TO axis. We apply the correction to the final TO measurement.

\subsection{Atom Interferometer}
\label{sec:aiapp}
Two beams drive stimulated Raman transitions between $|2S_{1/2},F{=}2,m_F{=}0\rangle$ and $|2S_{1/2},F{=}1,m_F{=}0\rangle$, counter-propagating along an axis within $5^\circ$ of $\hat{z}$, perpendicular to gravity. The Raman beams are detuned from the $|2S_{1/2}\rangle{\rightarrow} |2P_{3/2}\rangle$ transition by $-200$ MHz and drive fast $\pi/2$ pulses in 160 ns \cite{Cassella}. The difference in optical frequency between the two laser fields, labeled 1 and 2, for pulse $p$ is set near the ground-state hyperfine splitting $f_{12}^{(p)}{\approx}$803.5 MHz to stimulate the Raman transition. Each light field perturbs the ground states differently, which modifies the splitting from its bare resonance. Tuning $f_{12}^{(p)}$ to the perturbed state splitting improves the coupling rate. The third and fourth Raman pulses address the atoms with a modified laser frequency difference compared to the first two pulses $f_{12}^{(3,4)}{=}f_{12}^{(1,2)}{+}f_m$. The small modulation imprints an additional interferometer phase $2\pi f_m T$ that we use to tune the sensitivity of the interferometer (Fig. \ref{fig:decay}(d)). 

The Raman beams' polarizations follow a lin$\perp$lin scheme, so that one beam's polarization is approximately $(\sigma^+{+}\sigma^-)/\sqrt{2}$ and the other beam's is $(\sigma^+{-}\sigma^-)/\sqrt{2}$. Without orthogonalizing the beams' polarizations, multiple Raman pathways through Li's unresolved $|2P_{3/2},F'\rangle$ states would destructively interfere and preclude a transition for $m_F{=}0$. While allowing Raman transitions for $m_F{=}0$, lin$\perp$lin cannot drive Raman transitions for $m_F{=}{\pm} 2$, so those residual populations after OP cannot undergo an interferometer and do not contribute a systematic to the TO measurement.

The Stark-shifting beam must propagate as close to the $\hat{y}$ imaging axis as possible for its effect to be visible in the images, but it might retain a small projection onto $\hat{z}$ in our experiment. Any nonzero projection of the propagation axis onto $\hat{z}$ precludes a purely $\pi$-polarized beam. Because the TO axis and the $\hat{z}$ axis form a plane, it is possible for its polarization to be purely orthogonal to $\hat{z}$, corresponding to a linear combination of $\sigma^\pm$ components (close to $\hat{x}{=}(\sigma^+{+}\sigma^-)/\sqrt{2}$). We therefore use a polarization perpendicular to $\hat{z}$ for the TO measurement (see Appendix \ref{sec:polarization}). The beam is focused tightly to $w_z{\approx} 150$ ${\rm \mu m}$ along the interferometry axis and more weakly perpendicular to the interferometry axis, $w_x{\approx} 600$ ${\rm \mu m}$. 

\subsection{Signal scaling}
\label{sec:scalingapp}
While the overall scale factor of Eq. (\ref{eq:phase}) is not relevant for identifying the zero crossing at $\Delta_{\rm TO}$, the sensitivity of the measurement is proportional to the maximum phase difference. The maximum phase at fixed occurs centered in $x$ and on the sides of the beam profile where the intensity gradient is highest, at $z{=}{\pm} w_z/2$. The peak of the intensity gradient is proportional to the peak intensity $I_p{\propto} P/w_xw_z$ ($P$ is the optical power, here ${\approx}$3 mW) and gains another factor of $1/w_z$ when $I(x,z)$ is differentiated. This leads to a proportionality of the maximum phase difference at fixed $\Delta_L{\ne}\Delta_{\rm TO}$:
\begin{equation} 
|\Delta\phi|_{\Delta_L,{\rm max}}\propto PT\tau / w_xw_z^2. 
\label{eq:maxphase}
\end{equation}
The maximum interferometer phase difference that the Stark-shifting pulse induces in this work is ${\sim}\pi/10$. It would appear advantageous to maximize $P$, $T$, and $\tau$, while minimizing $w_z$. In practice, the parameters must satisfy some constraints. 

The waist $w_z$ must remain large enough to satisfy two criteria. First, the phase pattern with spatial scale ${\approx}w_z$ must be readily observable given the spatial resolution of the imaging system, here 13 ${\rm \mu m}$ per camera pixel. Second, each atom enters the interferometer with a randomly oriented thermal speed 
and covers a distance $z_{th}{=}(2T+T')v_{th}{\approx}(2T+\tau)v_{th}$ over the course of the interferometer. This itineracy thermally dephases the pattern unless $w_z{\gg} z_{th}$. 
$T$ must be small enough that the arm separation $\Delta z$ samples the intensity gradient finely enough to observe its spatial variation. 

The usable pulse power $P$ is limited by the atoms' incoherent response, single-photon scattering. Although the AC Stark shift and polarizability vanish at $\Delta_{\rm TO}$, single-photon scattering events from each of the $|2S_{1/2}\rangle{\rightarrow} |2P_{1/2}\rangle$ and $|2S_{1/2}\rangle{\rightarrow} |2P_{3/2}\rangle$ transitions still occur proportional to $P\tau$ (though inversely proportional to the square of the detuning from each transition). Such scattering events destroy the coherence of the interferometer arms, so this limits the product $P\tau$. In practice, we roughly optimized for target parameters using numerical simulations prior to setup.

\section{Data and image processing}
\label{sec:analysis}
To a first approximation, images of laser-cooled atomic densities exhibit two-dimensional Gaussian profiles. If the Stark-shifting beam is centered on the atomic density profile, any gradient at the center of the density distribution would be attributable to phase patterning from the Stark laser. In practice, experimental noise complicates this ideal situation; the atomic density distribution is not precisely Gaussian, the number of atoms fluctuates and drifts, and the position of the cloud fluctuates and drifts on a length scale comparable to the Stark beam size. These position offsets between the beam and underlying density profile mimic the signal. Furthermore, a low signal-to-noise ratio in the images makes it difficult to identify the signal in any individual image (see Fig. \ref{fig:analysis}(d)). These realities prohibit simply fitting the atomic density distribution to a Gaussian function and analyzing the residuals for the Stark signature. We instead develop an image analysis method that averages out fluctuations and drifts.

Written explicitly for each pulse state ${\in}\{0,1\}$ and sensitivity $\pm$, the four image types are:
\begin{equation}
\begin{array}{ll}
&{\bf i}_{1,j}^+(\Delta_{L,j})={\bf A}_j+{\bf B}_j+{\bf N}_j+{\bf S}_j(\Delta_{L,j})+{\bf Z}_j(\Delta_{L,j}) \\
&{\bf i}_{0,k}^+={\bf A}_k+{\bf B}_k+{\bf N}_k \\
&{\bf i}_{1,l}^-(\Delta_{L,l})={\bf A}_l-{\bf B}_l+{\bf N}_l-{\bf S}_l(\Delta_{L,l})+{\bf Z}_l(\Delta_{L,l}) \\
&{\bf i}_{0,m}^-={\bf A}_m-{\bf B}_m+{\bf N}_m. \\
\end{array}
\label{eq:I}
\end{equation}
For each shot, ${\bf A}$ is the atomic density profile, ${\bf B}$ describes background population gradients introduced by the interferometer, ${\bf S}(\Delta_L)$ is the Stark signal, ${\bf Z}(\Delta_L)$ comes from single-photon scattering, and ${\bf N}$ is imaging noise. 
A total of $\sim$330,000 images contribute to the measurement of $\Delta_{\rm TO}$.

We average each image type within a subset $s$. The average atomic density, background interferometer signal, and imaging noise are independent of image type, so they cancel in the residual images for $s$. 
\begin{equation}
\begin{array}{ll}
{\bf R}_s^+(\Delta_L)&=\langle{\bf i}_{1,j\in s}^+|_{\Delta_L}\rangle-\langle{\bf i}_{0,k\in s}^+\rangle \\ &\approx{\bf S}(\Delta_L)+{\bf Z}(\Delta_L). \\
{\bf R}_s^-(\Delta_L)&=\langle{\bf i}_{1,l\in s}^-|_{\Delta_L}\rangle-\langle{\bf i}_{0,m\in s}^-\rangle \\ &\approx-{\bf S}(\Delta_L)+{\bf Z}(\Delta_L).
\end{array}
\end{equation}
Subsets of 1,000 images, spanning 30 minutes of integration, produce the smallest uncertainty in our experiment.

The difference of residuals removes the effect of single-photon scattering induced by the Stark-shifting beam and provides direct access to the signal of interest.
\begin{equation}
{\bf D}_s(\Delta_L)={\bf R}_s^+(\Delta_L)-{\bf R}_s^-(\Delta_L)\approx 2{\bf S}(\Delta_L).
\label{eq:D}
\end{equation}
Averaging over all the subsets generates the insets shown in Fig. \ref{fig:DIms}: ${\bf D}(\Delta_L){=}\langle{\bf D}_s(\Delta_L)\rangle$.

Fitting the images for the TO frequency requires a fit for the dipole pattern. ${\bf D}$ expresses the Stark pattern most strongly when the Stark laser is tuned furthest from TO to $\Delta_M$ (the left-most point in Fig. \ref{fig:DIms}), 
about 400 MHz from $\Delta_{\rm TO}$. Following Eq. (\ref{eq:phase}), we fit ${\bf D}(\Delta_M)$ to a model 
\begin{equation}
\begin{array}{ll}
{\bf M}&(a,\theta,x_0,z_0,\sigma_{\tilde{x}},\sigma_{\tilde{z}})\\
&=a \tilde{z} \exp{(-(\tilde{z}-z_0)^2/2\sigma^2_{\tilde{z}})} \exp{(-(\tilde{x}-x_0)^2/2\sigma^2_{\tilde{x}})}, 
\end{array}
\end{equation}
where the image coordinates $(x,z)$ are rotated by an angle $\theta$ (the angle of the Raman beams to $\hat{z}$) about $(x_0,z_0)$ to $(\tilde{x},\tilde{z})$. The amplitude $a$, angle $\theta$, center $(x_0,z_0)$, and widths $\sigma_{\tilde{x}}$ and $\sigma_{\tilde{z}}$ are free parameters in the fit. 

The projection of ${\bf D}_s(\Delta_L)$ onto ${\bf M}$ quantifies the strength and sign of the dipole pattern as a function of the laser frequency.
\begin{equation}
P_s(\Delta_L)={\bf D}_s(\Delta_L)\cdot{\bf M}=\sum_{x,z} D_s(\Delta_L,x,z)M(x,z).
\end{equation}
The zero crossing of a fit to $P_s(\Delta_L)$ provides an estimate for $\Delta_{\rm TO}$ for each of 320 subsets, $\Delta_{{\rm TO},s}$ (see Fig. \ref{fig:DIms}). The result of this two-dimensional analysis is an average over subsets $\Delta_{\rm TO}{=}\langle\Delta_{{\rm TO},s}\rangle$, with a statistical uncertainty given by the standard error among the measurements of each subset: 3335.7(1.2) MHz. 
Simulating noisy fake data offers an opportunity to set a known $\Delta_{\rm TO}$ and check for extra systematic effects in the image analysis protocol, though we find none.

\section{Systematic effects}
\label{sec:sysapp}

\subsection{Decoherence of $F{=}1$ interferometer}
\label{sec:decay}
\begin{figure}[t!]
\centering
\includegraphics[width=0.47\textwidth]{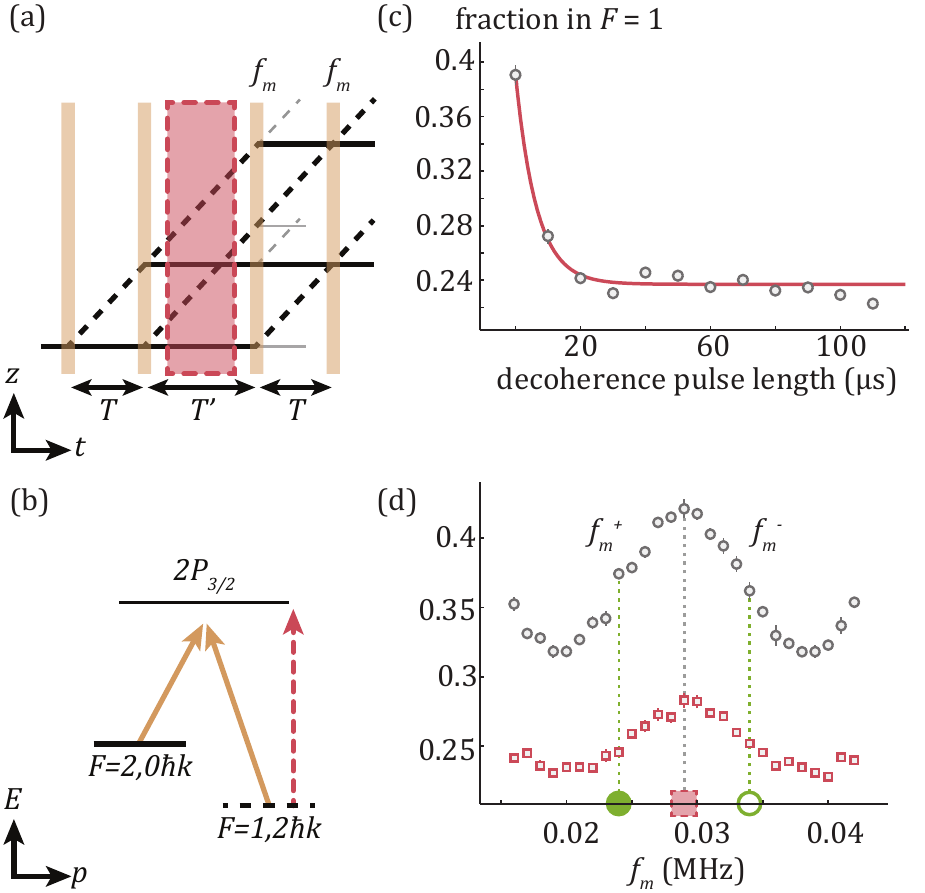}
\caption{\label{fig:decay} Decoherence of the complementary interferometer. (a) The four $\pi$/2 Raman pulses (gold rectangles) can drive two separate interferometers. We decohere the complementary (upper) interferometer by selectively addressing the $|2S_{1/2},F{=}1\rangle$ state with MOT repump light (dashed red pulse). Light gray lines indicate non-interfering beam splitter outputs. (b) Raman beams (solid gold) couple the interfering states. MOT repump light decoheres only the $F{=}1$ state. (c) Fixing the modulation frequency at the top of the fringe (dashed red square in (d)), the MOT repump pulse destroys the contrast of the complementary interferometer. (d) Population fringes without the decoherence pulse (gray circle data) represent the sum of the complementary interferometers' signals, which have identical phase. After the decoherence pulse, only contrast from the desired lower interferometer remains (open red square data). The dashed and filled red square indicates the fixed $f_m$ at which the decay curve in (c) was taken, while the solid (open) green circle indicates the $f_m^\pm$ used for positive (negative) phase sensitivity (see Fig. \ref{fig:analysis}).}
\end{figure}

After the second $\pi/2$ Raman pulse in the interferometer, each pair of components in a particular hyperfine state can close an interferometer with the two remaining pulses. Furthermore, their TOs differ by roughly the ground-state hyperfine splitting ${\sim}800\,$MHz, so this state impurity can introduce a substantial systematic shift. To restrict our measurement to the TO of the $|F{=}2,m_F{=}0\rangle$ state, a 70-${\rm \mu s}$ pulse of MOT repump light destroys the coherence of the $|F{=}1,m_F{=}0\rangle$ interferometer during $T'$ (Fig. \ref{fig:decay}) by driving single-photon scattering events. Without the decoherence pulse, each of the complementary interferometers contributes amplitude in the detected $F{=}1$ interferometer output and their signals add \cite{Cassella}. The peak output of the fringe is set by the sum of contrasts for each of the interferometers. We probe the contrast decay of the complementary interferometer by fixing $f_m$ at the top of the fringe (Fig. \ref{fig:decay}(d)) and scanning the duration of the decoherence pulse. The contrast of the complementary interferometer decays with a time constant of 6.5(1.7) ${\rm \mu s}$ (Fig. \ref{fig:decay}(c)), leaving behind a population fraction of $2{\times}10^{-5}$. Assuming half the population undergoes each of the interferometers, this systematic totals to ${\approx}(1{\times}10^{-5})800\,$MHz=0.02 MHz.

\subsection{Spectroscopy Zeeman shift}
\label{sec:specZeeman}
Because OP leaves atoms skewed towards the $m_F{=}{-}2$ state, the measurement of $\langle f_0\rangle$ is subject to a Zeeman shift proportional to the population asymmetry between $m_F{=}{\pm} 2$. The Zeeman shift for the $|2S_{1/2},F{=}2,m_F{=}{\pm} 2\rangle{\rightarrow} |2P_{1/2},F'{=}2,m_F'{=}{\pm} 1\rangle$ transition is roughly $\pm$0.760 MHz. With a population asymmetry of 12\% and roughly half the optical power contributing to these transitions due to the polarization, we estimate this to systematically shift the peak center by $0.09(2)$ MHz. 

Note that this Zeeman shift manifests nearly identically in both CAS$_z$ and CAS$_{\rm TO}$, since the polarizations are nearly identical along both axes and they are taken at the same magnetic field. Given that the Zeeman effect produces equal shifts in both, their relative comparison reveals the Doppler shift without any additional shift due to the magnetic field (Appendix \ref{sec:spec}).

The Stark-shifting laser's $\sigma^\pm$ polarization couples the $m_F{=}0$ ground state to $m_F'{=}{\pm} 1$ for the TO measurement. We note that while the transition energies for the two polarization components experience opposite Zeeman shifts up to the MHz level, the shifts to the coupling strengths 
effectively average out across the two components and the shift to TO remains negligible.

\subsection{Polarization}
\label{sec:polarization}
We calculate from theoretical matrix elements and experimental transition energies that TO varies maximally by 47 MHz \cite{Kien,Krutzik,MariannaLi,Sansonetti} between $\pi$ and $\sigma^\pm$ polarizations for $|F{=}2,m_F{=}0\rangle$. The polarization of the Stark-shifting laser along the TO axis must therefore be controlled to within several degrees. Since we cannot achieve pure $\pi$ polarization in our geometry (see Appendix \ref{sec:aiapp}), we aim for a pure linear combination of $\sigma^\pm$, which produces a maximal tensor shift that pushes the measured TO detuning upward. 

A Wollaston prism purifies the Stark beam's polarization with an extinction ratio of ${\sim} 10^5$. The purified beam passes sequentially through a $\lambda/2$ wave plate tilted to an angle $\theta_{\lambda/2}$ and a $\lambda/4$ wave plate tilted to $\theta_{\lambda/4}$ (Thorlabs WPH05M-670 and WPQ05M-670, respectively). Each is mounted on its own motorized rotation stage (Thorlabs PRM1Z8). After the wave plates, the beam encounters two in-plane broadband dielectric mirrors and one periscoping metallic mirror before passing through a vacuum window and impinging on the atoms. We probe the polarization of the Stark beam via two methods. 

First and more coarsely, we sample the polarization of the beam before the vacuum chamber with a polarizing beam splitter (PBS) and rotate the motorized wave plates to generate a polarization outside the chamber that closely matches the target polarization parallel to $\hat{x}$ (i.e. parallel to the plane of the optical table). This polarization occurs at $\theta_{{\rm PBS},\lambda/2}{=}6^\circ$ and $\theta_{{\rm PBS},\lambda/4}{=}4^\circ$. We perform a TO measurement at this polarization and at a series of linear polarizations incremented by $20^\circ$ (see Fig. \ref{fig:tensor}). Steps of $10^\circ$ in $\theta_{\lambda/2}$ rotate the polarization by $20^\circ$, so $\theta_{\lambda/4}$ requires steps of $20^\circ$ to follow. The wave plate angles therefore respect a fixed relationship of $(\theta_{\lambda/2}{-}6^\circ){=}2(\theta_{\lambda/4}{-}4^\circ)$ in Fig. \ref{fig:tensor}. These measurements trace out the polarization-dependent tensor variation of the polarizability. A fit to these data accesses two important features: the amplitude of the variation, as well as the central polarization, where $\Delta_{\rm TO}$ is maximized and the polarization most closely resembles $\hat{x}$. While theory predicts a $47$-MHz variation, we observe a variation of $56.9(4.7)$ MHz. Because some slight ellipticity may be present at the atoms using this method, the full tensor variation may be marginally larger. The degree of the ellipticity present using this method should produce a systematic much smaller than the uncertainty in the amplitude of the fit. 
The central polarization fits to $\theta_{0,\lambda/2}{=}12.4(9)^\circ$, corresponding to $\theta_{0,\lambda/4}{=}17(2)^\circ$. Vacuum windows can induce polarization rotations significant enough to shift the TO measurement through their birefringence \cite{SackettMeasurement,birefringence}. Therefore, this method of setting the polarization with a polarizer outside the vacuum chamber alone is likely insufficient, so we devise a separate method for better accuracy.

\begin{figure}[t]
\centering
\includegraphics[width=0.47\textwidth]{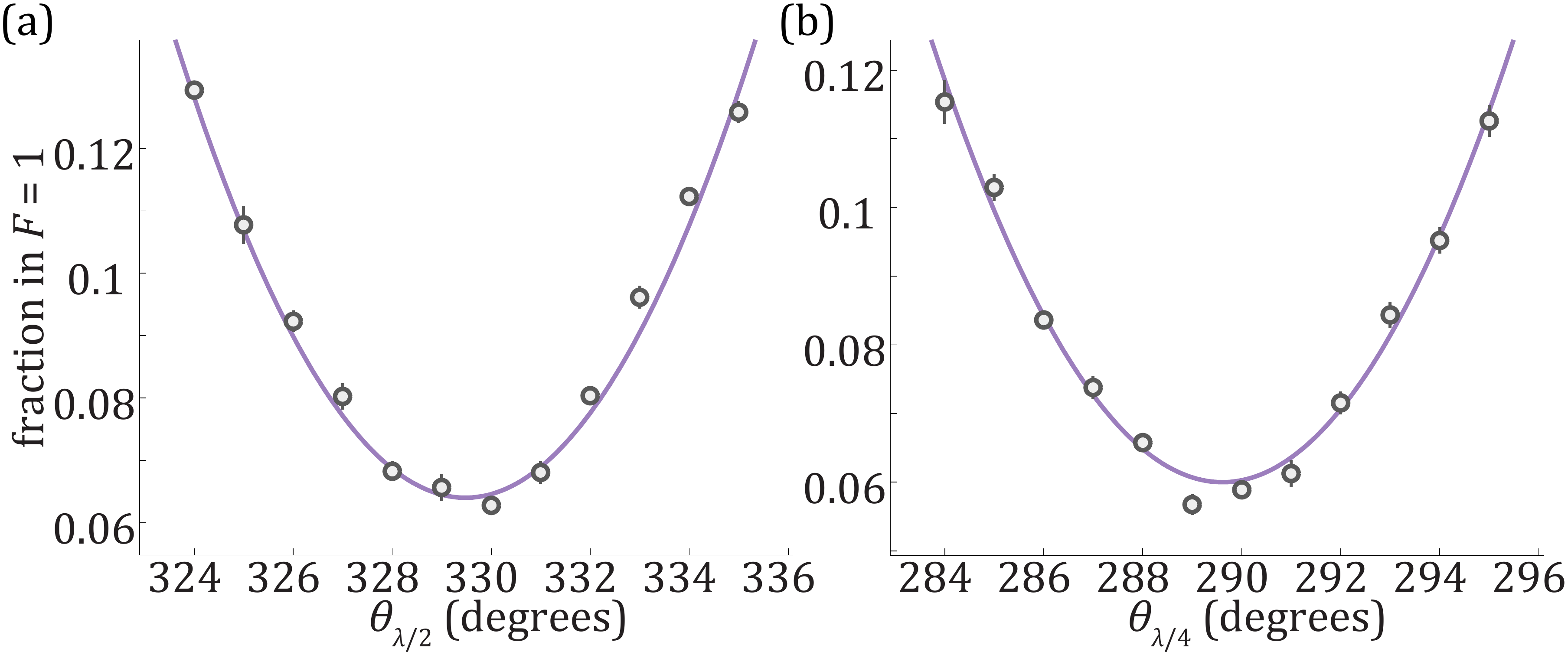}
\caption{\label{fig:wp} Wave plate optimization. When propagating along the Stark beam axis, the Stark beam's polarization can be precisely tuned to minimize scattering close to the $\pi$ polarization required for OP. This requires co-optimization of (a) the $\lambda/2$ angle and (b) the $\lambda/2$ angle. Fits for the minima are shown in purple. For these experiments, the laser is locked to $\Delta f^{D1,22}_{\rm TO}$ (see Fig. \ref{fig:cas}). The minima most closely identify $\pi$ polarization and we rotate the polarization 90$^\circ$ from there for the primary TO measurement.}
\end{figure}

Second, we probe the polarization \emph{in situ} by driving the OP transition with the Stark laser along the TO axis. Though the geometry does not allow for the Stark beam to be purely $\pi$-polarized as is required for OP, optimization of the polarization to this axis reduces the scattering by a factor of ${\sim}5$. We find the wave plate settings that minimize scattering on the OP transition at $\theta_{{\rm OP},\lambda/2}{=}329.5(1)^\circ$ and $\theta_{{\rm OP},\lambda/4}{=}289.6(1)^\circ$ (see Fig. \ref{fig:wp}). The polarization must rotate from there by $90^\circ$, so we rotate the $\theta_{\lambda/2}$ by $45^\circ$ and the $\theta_{\lambda/4}$ by $90^\circ$. The motorized rotation mounts specify a rotation accuracy of $0.2^\circ$, so we now assume each wave plate to be within this specification from the setting to optimally achieve a linear combination of $\sigma^\pm$ polarizations. These optimized polarizations are at $\theta_{{\rm TO},\lambda/2}{=}14.5^\circ$ and $\theta_{{\rm TO},\lambda/4}{=}19.6^\circ$. We expect this optimization to produce the more accurate result and use it for the precision measurement campaign. 

The discrepancy between the two polarization optimizations above provides a natural scale for the error in the wave plate settings. The methods disagree by $\Delta\theta_{\lambda/2}{=}\theta_{{\rm TO},\lambda/2}{-}\theta_{0,\lambda/2}{=}2.1(9)^\circ$ (a linear polarization uncertainty of $4(2)^\circ$), and by $\Delta\theta_{\lambda/4}{=}\theta_{{\rm TO},\lambda/4}{-}\theta_{0,\lambda/4}{=}3(2)^\circ$. The linear error scales the full tensor shift by the projection of the polarization onto the wrong axis. We use our measured tensor shift to determine the uncertainty: $\sin^2(4^\circ)(57$ MHz$){=}0.3$ MHz. The ellipticity introduced by the $\lambda/4$ wave plate can only vary the polarization between a pure linear combination of $\sigma^\pm$ and, at worst, an equal superposition of $\pi$ and the circular components. The maximal projection of half the power onto $\pi$ renders this systematic only half as potent as that from the linear polarization angle: $0.5\sin^2(3^\circ)(57$ MHz$){=}0.1$ MHz. We do not investigate the presence of or correct for polarization drift as in reference \cite{SackettMeasurement}.

The final value of $\Delta_{\rm TO}$ uses data obtained only with the second method of polarization control (see Figures \ref{fig:stats} and \ref{fig:oned}). It does not include the data presented in Fig. \ref{fig:tensor}. Correcting for the polarization-dependent tensor shifts at each data point therein requires accurately knowing the full size of the tensor shift. Given the ambiguity between theory and our results, we omit those data from our final analysis and sacrifice the uncertainty reduction they offer.

\subsection{Broadband laser emission}
\label{sec:ase}
\begin{figure}[h]
\centering
\includegraphics[width=0.42\textwidth]{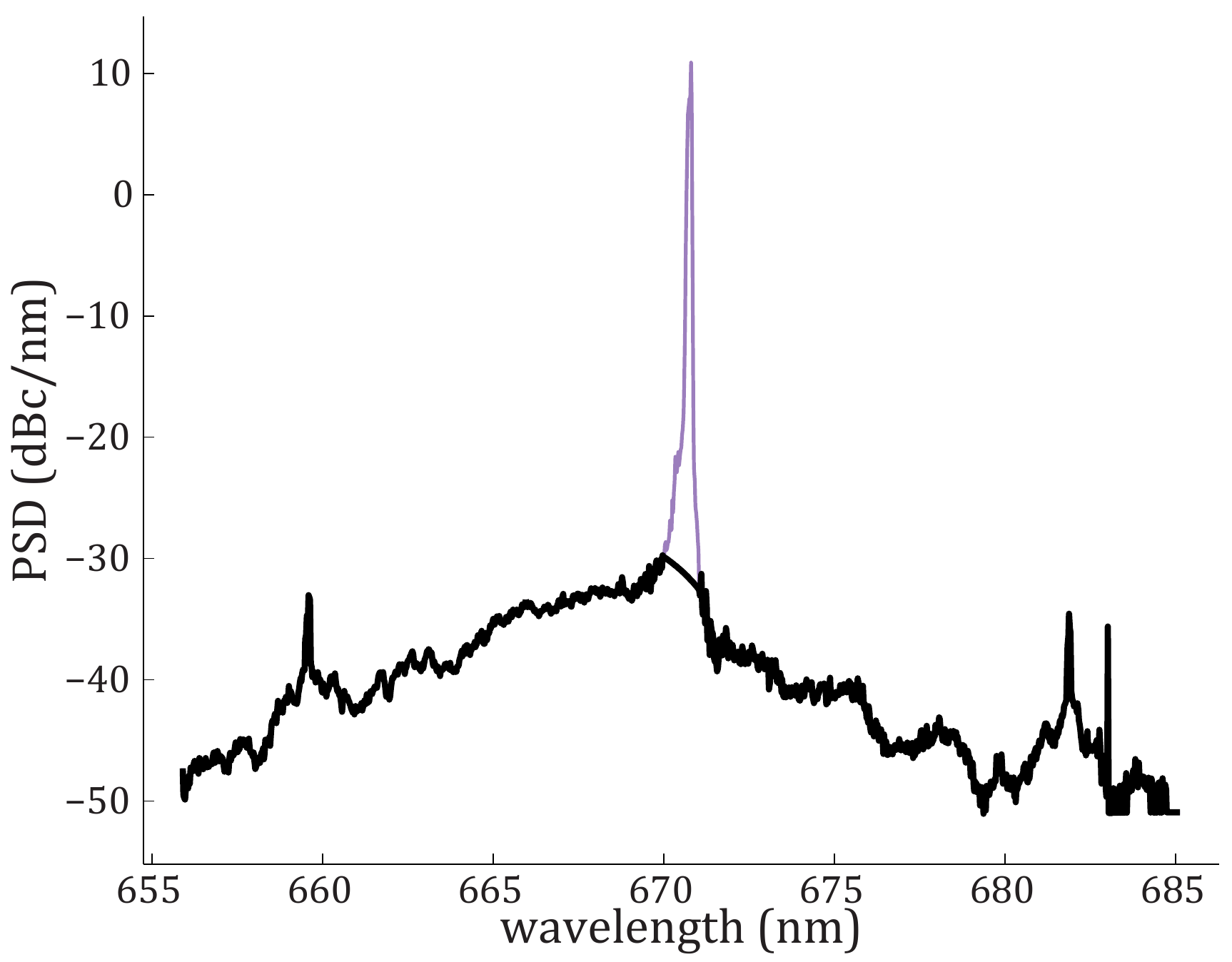}
\caption{\label{fig:psd} Power Spectral Density of the Stark laser's Amplified Spontaneous Emission. To calculate the associated systematic shift, we truncate the lasing peak (thin purple trace) and restrict attention to the remaining broadband emission (thick black trace). This spectrum is imaged at a resolution bandwidth of 0.022 nm, though the noise bandwidth is larger.}
\end{figure}

Diode lasers output Amplified Spontaneous Emission (ASE), a broadband spectrum spanning ${\sim}100\,$ nm. 
We set the Stark laser near TO and record its power spectral density (PSD) using a grating spectrometer (Princeton Instruments Acton SpectraPro SP-2300 with PIXIS 400 CCD). Imperfect alignment in the spectrometer imaging system asymmetrically distributes photons from the lasing peak across a series of pixels. We identify the associated artifacts using features common to the peaks of a neon-argon calibration lamp. The artifacts span 1 nm, so we ignore spectral information within ${\pm}$0.5 nm of the lasing peak (see Fig. \ref{fig:psd}). We sum the atom interferometer phases from a monochromatic lasing peak and the ASE separately and solve for the zero crossing of the phase shift. This computed shift in TO totals $<$0.1 MHz. Uncertainty in this shift derives from the wavelength calibration of the spectrum, the total power in the lasing peak used to calculate the PSD in dBc, and the parameters used for truncating the lasing peak as shown in Fig. \ref{fig:psd}. 

The phase-locked laser beat note provides spectral information in a more narrow-band region closer to the laser peak. 
The Gaussian distribution near the peak of the beat note is symmetric and exhibits -3-dB points separated by 1.4 MHz. No asymmetry is apparent in the Lorentzian tails. 
We conclude that only the broadband emission in Fig. \ref{fig:psd} systematically shifts the TO measurement.

\subsection{Hyperpolarizability}
TO measurements benefit from the Stark shift zero crossing being independent of intensity, which is hard to calibrate \emph{in situ}. There is a higher-order shift from the hyperpolarizability, proportional to the square of the intensity. Fourth order in perturbation theory, this term involves sums over four-photon processes. A three-level model with a ground state and two excited states (e.g. $|g\rangle{=}|2S_{1/2}\rangle$, $|a\rangle{=}|2P_{1/2}\rangle$, and $|b\rangle{=}|2P_{3/2}\rangle$) is the minimal model capable of cataloging all the dominant four-photon couplings ($g{\rightarrow} a{\rightarrow} g{\rightarrow} a{\rightarrow} g$; $g{\rightarrow} a{\rightarrow} g{\rightarrow} b{\rightarrow} g$; $g{\rightarrow} b{\rightarrow} g{\rightarrow} a{\rightarrow} g$; and $g{\rightarrow} b{\rightarrow} g{\rightarrow} b{\rightarrow} g$). 

We compute the energy eigenvalues of the three-level Hamiltonian with a drive detuned between the two excited-state transitions. The perturbed energy of $|g\rangle$ can be expanded in a power series of the drive intensity $I$, where term ${\propto}I$ is the polarizability and the term ${\propto}I^2$ is the hyperpolarizability. The hyperpolarizability's zero crossing coincides with the polarizability's zero crossing at TO, so it contributes no significant shift to TO at our level of precision.

\subsection{Single-photon scattering}
\label{sec:scattering}

\begin{figure}
\includegraphics[width=0.42\textwidth]{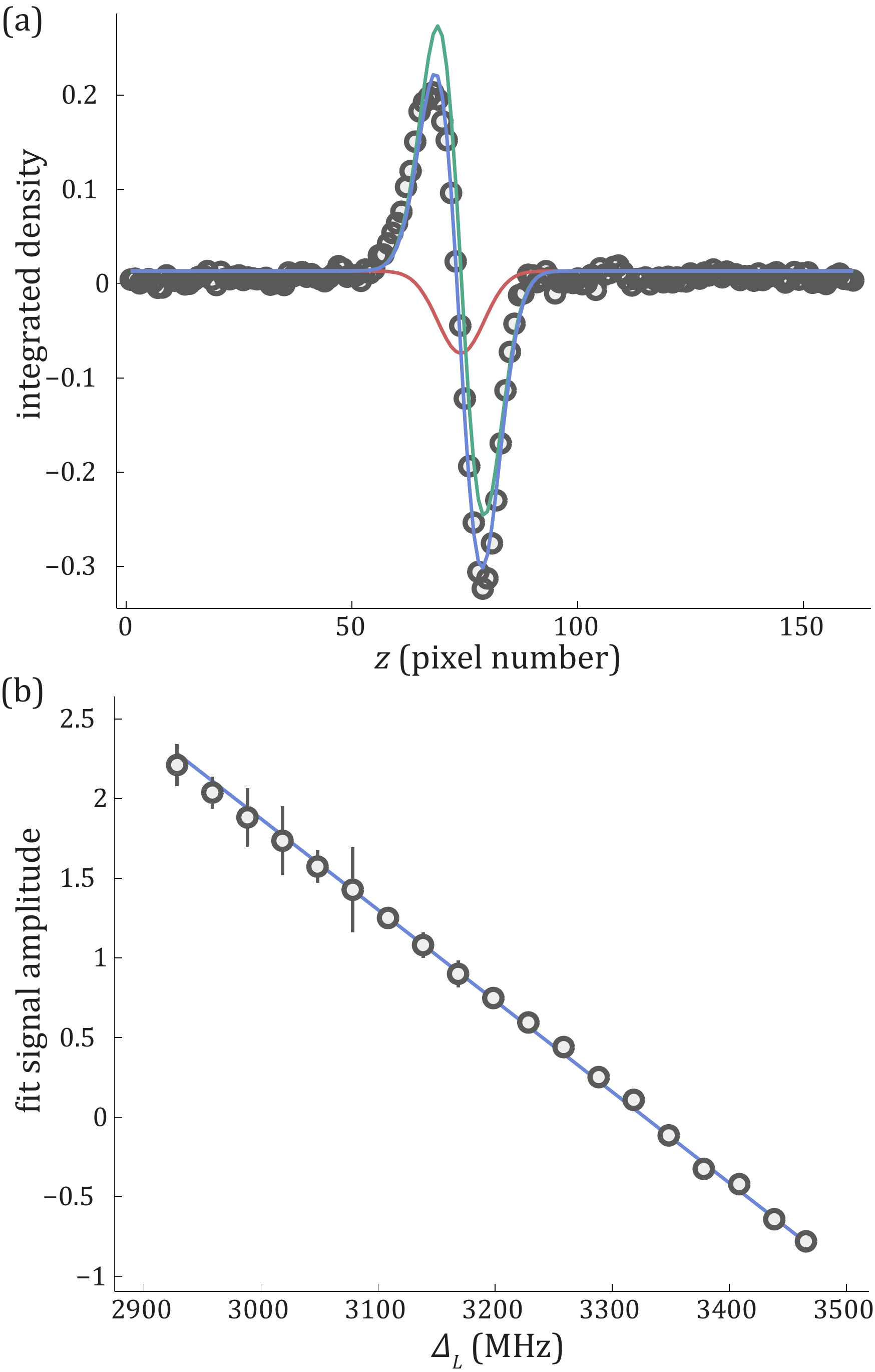}
\caption{\label{fig:oned} One-dimensional analysis of scattering systematic. (a) After rotating out the tilt of ${\bf D}(\Delta_L)$ (here shown for $\Delta_M$), we integrate along $x$ and rescale the values to the local atom density. We fit a full model (blue line) as the sum of an odd signal (green line) and even scattering peak (red line). That the scattering peak is negative implies that scattering expresses more strongly in ${\bf R}^-$ (b) Amplitudes of the signal portion of the fit in cross through zero. The blue line shows a fit to the amplitudes, which crosses zero at $\Delta_{\rm TO}$.}
\end{figure}

The analysis presented in Appendix \ref{sec:analysis} is not impervious to the monopole-shaped pattern generated by single-photon scattering. It assumes that the scattering pattern is identical for each phase sensitivity. Admitting a small mismatch between the sensitivities spoils this assumption and introduces an asymmetry that prevents perfect cancellation in ${\bf D}(\Delta_L)$. That is, a small component of ${\bf Z}_s(\Delta_L)$ may survive in Eq. (\ref{eq:firstD}) or (\ref{eq:D}). There is still no shift as long as the scattering pattern is perfectly centered on the dipole signal pattern. If so, the projection of the scattering monopole onto ${\bf M}$ is 0 because it involves integrating the product of an odd function and an even function with the same center. Any systematic position offset between the center of the scattering pattern and the center of the signal pattern introduces a systematic offset in the projections $P_s(\Delta_L)$.

Though the scattering and signal patterns originate from the same Stark beam profile, an offset between the patterns may still arise. While the dipole signal pattern remains stationary, atoms that scatter a photon do recoil at 8.5 cm/s. There is an angle $\lesssim 5^\circ$ between the Stark beam and the imaging beam. During the ${\sim}$100 ${\rm \mu}$s between scattering in the middle of the interferometer and detection, the monopole pattern can drift in the imaging plane by $\sin(5^\circ)$(8.5 cm/s)(100 ${\rm \mu}$s) ${\sim}1$ ${\rm \mu}$m and produce a systematic shift to the analysis in Section \ref{sec:analysis}.

The phase-patterned signal specific to the work presented here contains information relevant to TO only along one axis, $\hat{z}$. Rotating out the $4.41^\circ$ tilt apparent in ${\bf D}(\Delta_L)$ and integrating along $x$ reduces the signals to one dimension (see Fig. \ref{fig:oned}(a)). Rescaling to the local atomic density from a similar one-dimensional integration of the averaged unpulsed shots removes any asymmetry introduced by an offset between the center of the Stark beam and the background ${\bf A}{\pm}{\bf B}$. We fit the resulting trace to a sum (blue) of the monopole scattering contribution (red) and the dipole signal contribution (green). Only the amplitude of the signal portion of the fit is germane to the TO measurement, so we plot those amplitudes as a function of Stark laser wavelength in Fig. \ref{fig:oned}(b) and fit for the zero crossing. This scattering-corrected result is shifted down from the result of the two-dimensional analysis (Fig. \ref{fig:stats}) by 7.76 MHz. The uncertainty in the fit combines the statistical uncertainty of the measurement and the systematic uncertainty from this scattering offset.

The small excited-state fine-structure splitting exacerbates the effect of scattering in Li, due to the relatively small detuning of TO from the resonances. While scattering introduces the largest systematic shift in this measurement, it would be a smaller concern in different atomic species or for the phase-patterning technique more generally.

\bibliographystyle{apsrev4-1}
\bibliography{main}

\end{document}